\documentclass[12pt]{iopart}

\usepackage{graphicx}
\usepackage{bm}
\usepackage{color,graphicx,latexsym,psfig}

\usepackage{iopams}

\newcommand{\vl}[1]{\textcolor{black}{#1}}

\begin{document}

\title[High-field ESR in CaCu$_2$O$_3$]{{High-field ESR studies of the quantum spin magnet CaCu$_2$O$_3$}}
\author{M Goiran$^{1}$, M Costes$^{1}$, J M Broto$^{1}$,
F C Chou$^{2}$\footnote{Present address: Center of Condensed Matter Sciences, National Taiwan University, Taipei 106, Taiwan} , R Klingeler$^{1,4}$,
E Arushanov$^{1,3,4}$, S-L Drechsler$^{4}$, B B\"uchner$^{4}$ and V Kataev$^{4}$}
\address{$^{1}$Laboratoire National des Champs Magn\'etiques Puls\'es, 31432 Toulouse Cedex 04, France}
\address{$^{2}$Center for Materials Science and Engineering, Massachusetts Institute of Technology, Cambridge, Massachusetts 02139}
\address{$^{3}$Institute of Applied Physics, Academy of Sciences of Moldova, MD 2028 Chisinau, Moldova}
\address{$^{4}$Leibniz Institute for Solid State and Materials Research IFW Dresden, P.O. Box 270116, D-01171 Dresden, Germany}
\eads{\mailto{goiran@lncmp.org} and \mailto{v.kataev@ifw-dresden.de}}

\date{\today}

\begin{abstract}

We report an electron spin resonance (ESR) study of the $s\!=\!1/2$-Heisenberg pseudo-ladder magnet CaCu$_2$O$_3$ in pulsed magnetic fields up to
40~T. At sub-Terahertz frequencies we observe an ESR signal originating from a small amount of uncompensated spins residing presumably at the
imperfections of the strongly antiferromagnetically correlated host spin lattice. The data give evidence that these few percent of ''extra'' spin
states are coupled strongly to the bulk spins and are involved in the antiferromagnetic ordering at $T_N\!=\!25$~K. By mapping the
frequency/resonance field diagram we have determined a small gap for magnetic excitations below $T_N$ of the order of {$\sim\!0.3\,-\,0.8$\,meV.}
Such a small value of the gap explains the occurrence of the spin-flop transition in CaCu$_2$O$_3$ at weak magnetic fields $\mu_0H_{sf}\!\sim\!3$~T.
Qualitative changes of the ESR response with increasing the field strength give indications that strong magnetic fields reduce the antiferromagnetic
correlations and may even suppress the long-range magnetic order in CaCu$_2$O$_3$. ESR data support scenarios with a significant role of the
''extra'' spin states for the properties of low-dimensional quantum magnets.

\end{abstract}

\submitto{\NJP}
\maketitle

\section{Introduction}

\begin{figure}
\begin{center}
\includegraphics[angle=0,width=0.6\columnwidth]{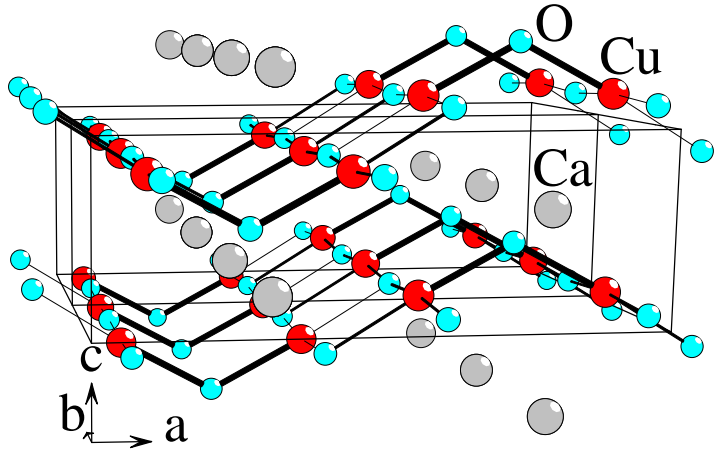}
\end{center}
\caption{{} The crystal structure of CaCu$_2$O$_3$. Cu-O chains running along the $b$ axis are coupled into two-leg ladders. Strongly
bent rung bonds in the ladders are visualized by thick lines.} \label{structure}
\end{figure}

Imperfections in quasi-one- or two-dimensional spin lattices \cite{richter} may result in peculiar changes of the properties of low-dimensional
quantum antiferromagnets. The host spin system around the defects such as nonmagnetic vacancies, doped ''extra'' spins or spin states at lattice
edges is strongly disturbed which is particulary pronounced in the Heisenberg magnets comprising small spins (e.g. $s\!=\!1/2$). A prominent example
is the doping of a two-dimensional (2D) $s\!=\!1/2$ Heisenberg antiferromagnet (HAF) with nonmagnetic impurities. Local magnetic moments associated
with nonmagnetic Zn dopants in the CuO$_2$ planes of high temperature superconductors have been observed by a number of experimental techniques (see
e.g.\ Ref.~\cite{Xiao90,Kataev90,Alloul91}) in accord with theoretical predictions \cite{Khaliullin97,Vojta00,Wang02,Hoglund}. Another interesting
aspect of imperfect spin lattices is the enhancement of antiferromagnetic correlations near impurities in low-D HAFs regardless the presence or
absence of long range order in the ground state of the host \cite{Bulut89,Martins97,Laukamp98}. Thus, long-range AF order may occur even in
spin-gapped compounds, likely as it happens in the Zn-doped two-leg $s\!=\!1/2$-ladder SrCu$_2$O$_3$ \cite{Azuma97,Ohsugi99}. Theory also predicts
that owing to the coupling of the impurity induced spin states with the host spin lattice the former may interact strongly via the AF  background
\cite{Chen00}. Experimentally one finds, that, for instance, Zn-doping induced spin moments in the CuO$_2$ planes of La$_{1-x}$Sr$_x$CuO$_4$ indeed
order antiferromagnetically below the N\'{e}el temperature of the Cu spin lattice \cite{Huecker02}. Recently, experimental indication of the strong
interaction between impurity and bulk spins in the $s\,=\,1/2$ low-D HAF CaCu$_2$O$_3$ have been reported by Kiryukhin {\it et al.}
\cite{Kiryukhin01}. CaCu$_2$O$_3$ has a crystal structure similar to the two-leg spin-ladder compound SrCu$_2$O$_3$ which exhibits a large spin gap
of about 420 K and is nonmagnetic at low $T$ \cite{Azuma94}. However, in contrast to the Sr counterpart where Cu spin chains parallel to the $b$ axis
are coupled in the $ab$ planes into ladders via a strong rung AF exchange, in CaCu$_2$O$_3$ the Cu-O-Cu bond angle in the rungs deviates
significantly from 180$^\circ$, resulting in a reduced rung coupling (figure~\ref{structure}). Remarkably, no hint for a spin gap in the Ca compound
was obtained so far. This suggests that the corrugation of the ladders changes the spin topology from nearly isolated two-leg ladders to
pseudo-ladders with significant interladder interactions, i.e.\ the coupled spin chains in this pseudo-ladder compound form anisotropic bilayers
parallel to the $bc$ plane \cite{Kiryukhin01,Kim03,Drechsler04,Sengupta04}. Owing to an appreciable inter-plane magnetic exchange along the $c$
direction $J_c$ concomitant with a strongly reduced rung interaction $J_r$ CaCu$_2$O$_3$ orders antiferromagnetically at $T_N\simeq 25$ K
\cite{Kiryukhin01,Ruck01}. Below 300 K the magnetic susceptibility $\chi$  of CaCu$_2$O$_3$ is quite small due to the strong intra-chain AF coupling
$J_b\!=\!J_{\!\parallel}\!\sim\!2000$ K. The observed {weak} Curie-like $T$ dependence of $\chi(\!T\!<\!300$ K) suggests that the magnetization in
this temperature range is determined by a small amount of {uncompensated} spins. {Surprisingly, a strong reduction of the susceptibility $\chi$ below
$T_N$ suggests that these spin states participate in the AF ordering of the host and therefore cannot be ascribed to some impurity phase}
\cite{Kiryukhin01}.

To get more insight into the unusual interplay between the ''extra'' spins and the host spin lattice, we have measured electron spin resonance (ESR)
of single crystals of CaCu$_2$O$_3$ in the sub-Teraherz frequency range in pulsed magnetic fields. ESR measurements which probe locally the magnetism
of paramagnetic centres show clearly that in fact these spins are involved in the magnetic ordering. The interplay with the host shows up in {the
critical broadening of the spectrum at $T\!\gtrsim\! T_N$, as well as in the reduction of the intensity of the signal and the shift of the resonance
field at $T\!<\!T_N$.} Remarkably, in fields above 20~T {these features are not present anymore:} The ESR signal  narrows and its intensity increases
{down to the lowest temperature} indicating the reduction of the AF fluctuations and of the long-range order. The field dependence of the ESR line
reveals a small gap of the order of  {$\sim\!0.3\,-\,0.8$\,meV} for {magnon} excitations in the ordered state which can explain the occurrence of a
spin-flop transition at a relatively small magnetic field of $\sim\!3$ T. \vl{ Our ESR data suggest that such ''extra'' states may interact strongly
with the host spin lattice in CaCu$_2$O$_3$ which may be also of relevance for the magnetic ordering of the other frustrated low-dimensional quantum
antiferromagnets. }

\section{Experiment and results}
\label{experiment}

CaCu$_2$O$_3$  crystallizes in the orthorhombic symmetry, space group Pmmn, with $a\!=\!9.949$~\AA,\ $\!b\!=\!4.078$~\AA\ and $\!c\!=\!3.460$~\AA\ at
$T\!=\!10$ K. Single crystals studied in the present work  have been grown using the travelling solvent floating zone  (TSFZ) method with CuO as a
flux. The details of the sample preparation and their characterization can be found in Ref.~\cite{Kiryukhin01}. \vl{The electron probe microanalysis
data taken at 10 points of the crystal surface of size $2\,\times\,2$\,mm$^2$ reveal the following composition of the elements. Ca: 0.854 (0.023),
Cu: 2.039 (0.056) and O: 3.005 (0.082). We note that an appreciable excess of Cu and a respective deficiency of Ca is quite typical for this material
\cite{Ruck01}.}

\begin{figure}
\begin{center}
\includegraphics[angle=-90,width=0.9\columnwidth]{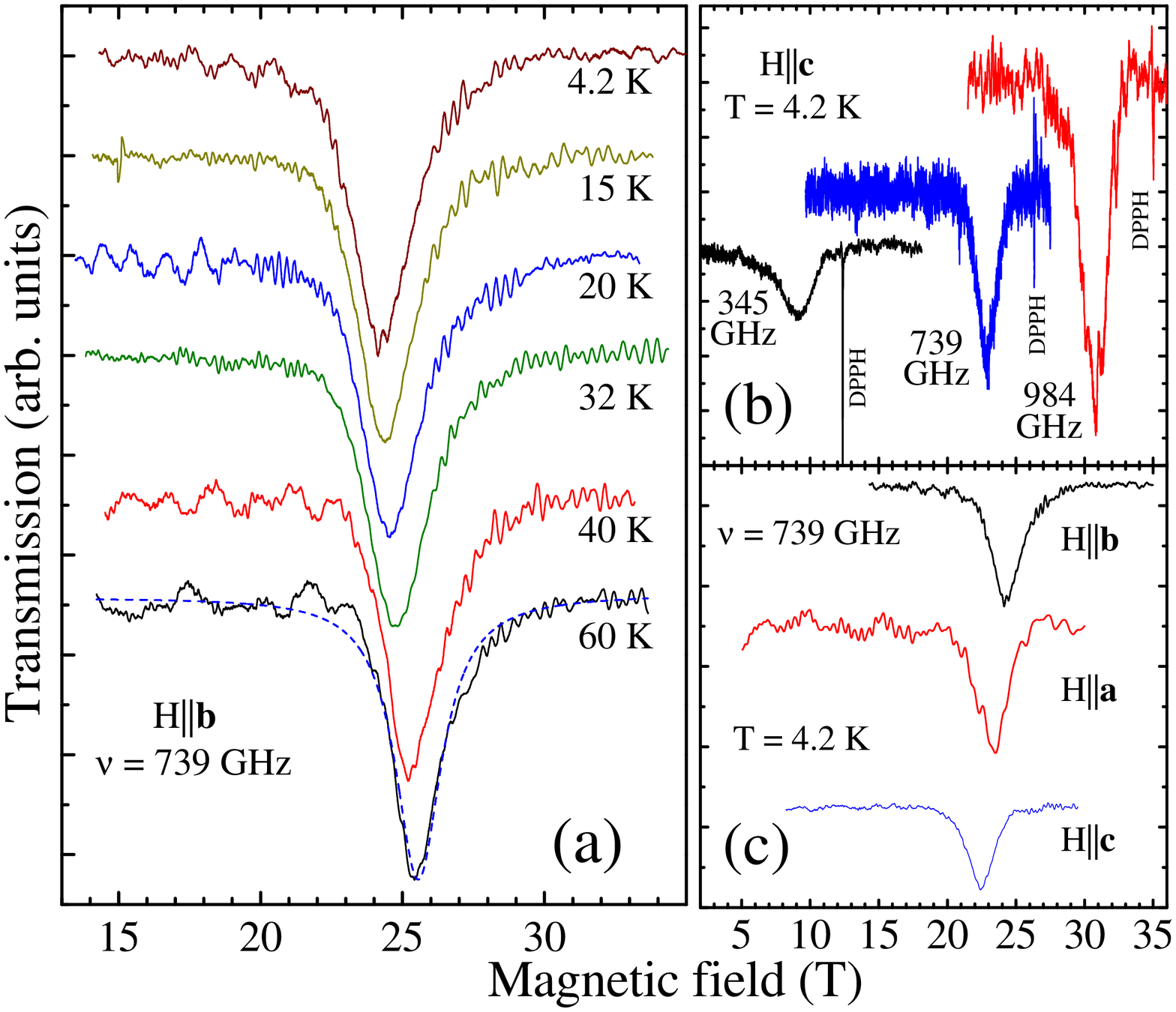}
\end{center}
\caption{{} Representative ESR spectra of CaCu$_2$O$_3$: (a) ESR signal at 739~GHz for $H\parallel b$~axis  at several selected
temperatures. The dashed curve is a representative Lorentzian fit of the signal at $T= 60$ K; (b) The ESR signal at different frequencies for
$H\!\parallel\!c$ axis at 4.2 K. Sharp resonances correspond to the {diphenyl-picryl-hydrazyl} (DPPH) field marker; (c) ESR spectra at 739~GHz and
4.2~K for $H$ applied along three crystallographic axes $a$, $b$ and $c$, respectively. } \label{spectra}
\end{figure}

\begin{figure}
\begin{center}
\includegraphics[angle=0,width=0.5\columnwidth]{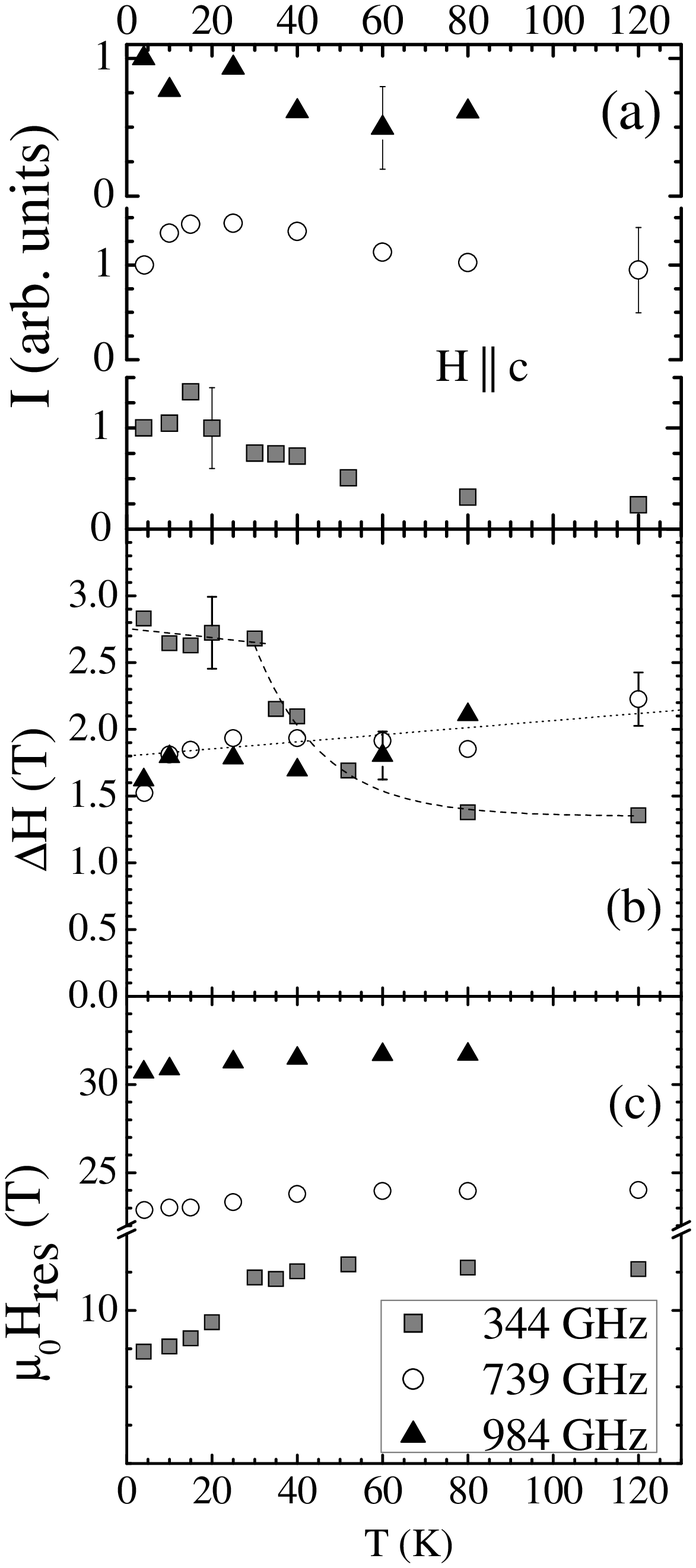}
\end{center}
\caption{Temperature dependence of the intensity $I(T)$ {normalized to the value at 4.2 K} (a), the width $\Delta H(T)$ (b) and the resonance field
$H_{res}(T)$ (c) of the ESR signal measured at frequencies 344 GHz, 739 GHz and 984 GHz, respectively, for the direction of the magnetic field
parallel to the $c$ axis. Dotted and dashed lines are guides for the eye. {The error bars in (c) are within the size of the symbols.}}
\label{parameters_c}
\end{figure}

The high field ESR experiments were performed up to 40~T in a resistive coil driven by a capacitor bank. The three
principal axes of the single crystal were successively set parallel to the magnetic field. The excitation energies were
provided by an optically pumped far infrared cavity and by a Gunn diode (96~GHz). Conventional ESR measurements in the
X-band were performed with a Bruker ESP 300 E Spectrometer. Magnetic susceptibility in the temperature range 2~K -
300~K and magnetization measurements up to 5~T were carried out with a Quantum Design Superconducting Quantum
Interference Device. Conventional ESR measurements at a frequency $\nu\!\simeq\!10$~GHz (X-Band) yield no signal and
those at $\nu\!\simeq\!100$~GHz reveal only a very weak resonance response which cannot be associated with strong
changes of bulk magnetic properties at low temperatures. In contrast, an order of magnitude stronger resonance
absorption has been observed at much higher frequencies and much stronger magnetic fields. Representative ESR spectra
of CaCu$_2$O$_3$ in the sub-Terahertz frequency domain are shown in figure~\ref{spectra}. The ESR spectrum consists of a
single line close to a Lorentzian shape with a slightly anisotropic resonance field $H_{res}$ typical for the resonance
response of the Cu$^{2+}$ ions with anisotropic $g$ factor (see below). By fitting the signal with a Lorentzian line
profile the intensity $I$, the resonance field $H_{res}$ and the width $\Delta H$ of the resonance have been
determined. The temperature dependence of these quantities is shown in figure~\ref{parameters_c} for the case of
$H\!\parallel\!c$ axis.

In pulsed magnetic field experiments the accurate determination of the intensity of the ESR line is difficult due to the poor signal-to-noise ratio
as compared to conventional ESR technique. Moreover, a large background due to the voltage induced in the detector by the field pulse has to be
subtracted from the total signal. Therefore the data shown in figure~\ref{parameters_c}(a) have large error bars and scatter considerably. Owing to
these technical problems the ESR intensities $I$ obtained at different frequencies cannot be directly compared and the $I(T)$ dependences in
figure~\ref{parameters_c}(a) at each frequency are normalized to a respective 4.2~K value. Fortunately, the linewidth $\Delta H$ and the resonance
field $H_{res}$ can be measured with a much better accuracy. One can recognize noticeable differences in the behaviour of these parameters vs.
temperature, depending on the excitation frequency and thus on the field range of the measurement. $\Delta H$  increases at the smallest ESR
frequency $\nu\!=\!344$ GHz appreciably in the temperature range 25~K $\lesssim\!T\!\lesssim\!60$~K and saturates below 25 K.
(figure~\ref{parameters_c}(b)). However, the ESR signal measured at higher frequencies and much stronger magnetic fields shows a continuous decrease
of $\Delta H$ in the whole temperature range. $H_{res}$ is practically constant at higher temperatures and decreases below $\sim\!25$ K, as can be
seen in figure~\ref{parameters_c}(c). Note that the shift of the resonance is most pronounced for the smallest ESR frequency and the field of the
measurement. This is illustrated in figure~\ref{shift} where the shift of $H_{res}$ at $T\!=\! 4.2$ K relative to its high temperature value is
plotted as a function of magnetic field $H$.

Remarkable changes of ESR at low $T$ are evident also in figure~\ref{branches} where the $\nu$ vs. $H_{res}$ diagram at a high (80 K) and at a low
temperature of 4.2 K is shown in the inset and in the main panel, respectively. {It is possible to fit {these} data} assuming a simple linear
frequency/field relation in a form $\nu\!=\!\Delta + (g\mu_B/h)H$, where $\Delta$ is the zero-field frequency offset, $g$ is the Lande $g$ factor,
$\mu_B$ is the Bohr magneton and $h$ is the Planck constant, respectively. The fit at $T\!=\! 80$ K yields $g$ factors amounting to 2.21 and 2.08 for
$H\!\parallel\!c$ and {{$b$} axes, respectively, which is usual for a Cu$^{2+}$ ion in a square-planar ligand coordination (figure~\ref{structure}).
The offset $\Delta$ is negligibly small, as expected for ESR of a $S\!=\!1/2$ paramagnetic ion having a Kramers degeneracy of the ground state. In
contrast, the fit at $T\!=\! 4.2$ K yields an appreciable offset $\Delta\!\simeq\!70$\,GHz corresponding to about 0.3 meV, or 3.4 K, which signals
the occurrence of an energy ({magnon}) gap at low {temperatures}.

\begin{figure}
\begin{center}
\includegraphics[angle=0,width=0.7\columnwidth]{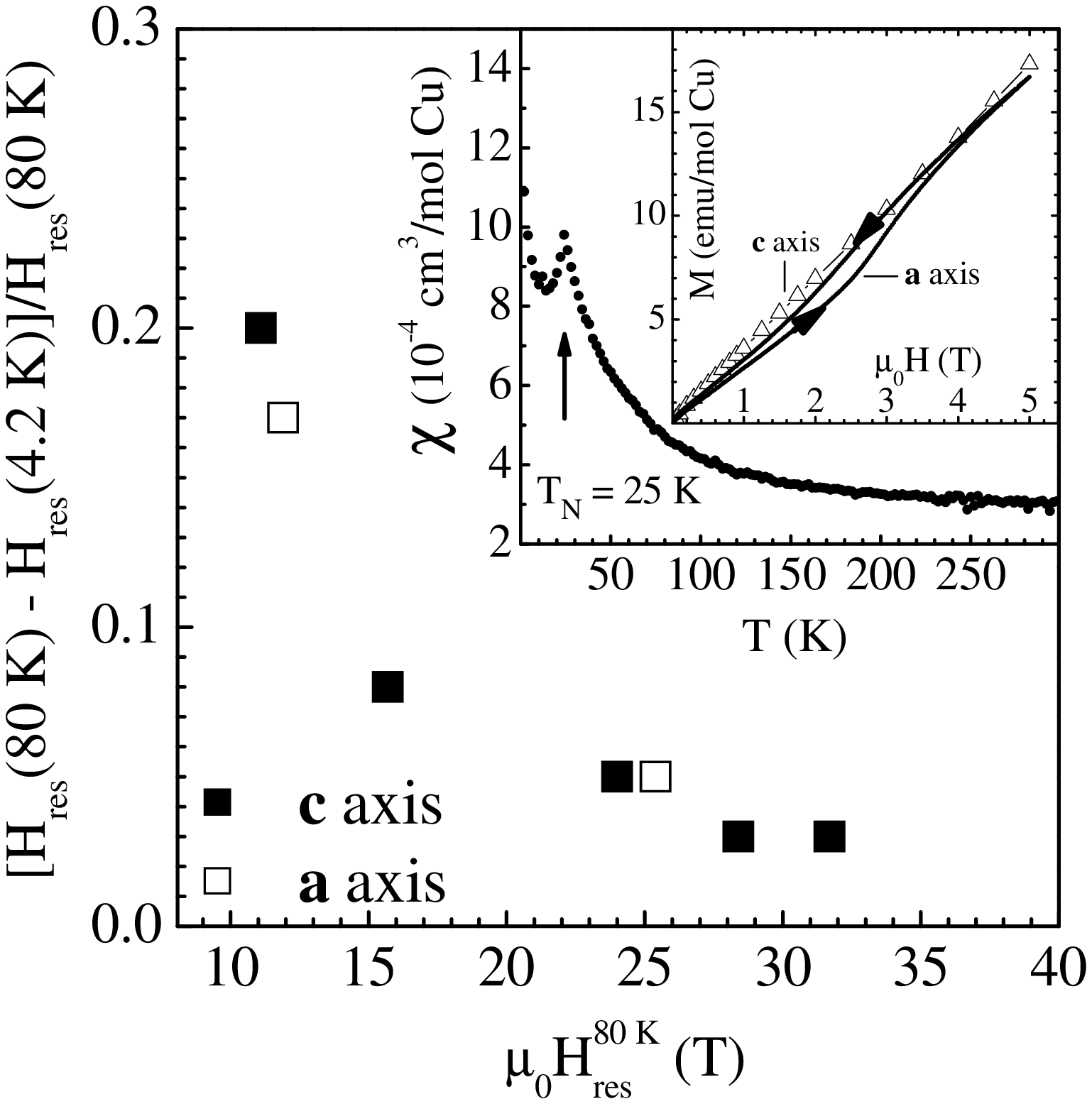}
\end{center}
\caption{The relative shift of the resonance field $H_{res}$ at $T\!=\!4.2$ K as a function of the magnetic field. Inset: Temperature dependence of
the static magnetic susceptibility $\chi(T)$ measured at $H\!=\!25$ Oe for $H\!\parallel\!a$ axis, and the field dependence of the magnetization
$M(H)$ for $H\parallel\!a$ and $c$ axes at $T\!=\!4.2$ K. Note the N\'eel peak in $\chi(T)$ indicating a transition to the AF ordered state at
$T_N\!=\!25$ K and a nonlinear hysteresis behaviour of $M(H)$ for $H\!\parallel\!a$ evidencing a field-induced {spin-flop} transition at
$\mu_0H_{sf}\!\sim\!3$ T.} \label{shift}
\end{figure}

\begin{figure}
\begin{center}
\includegraphics[angle=0,width=0.7\columnwidth]{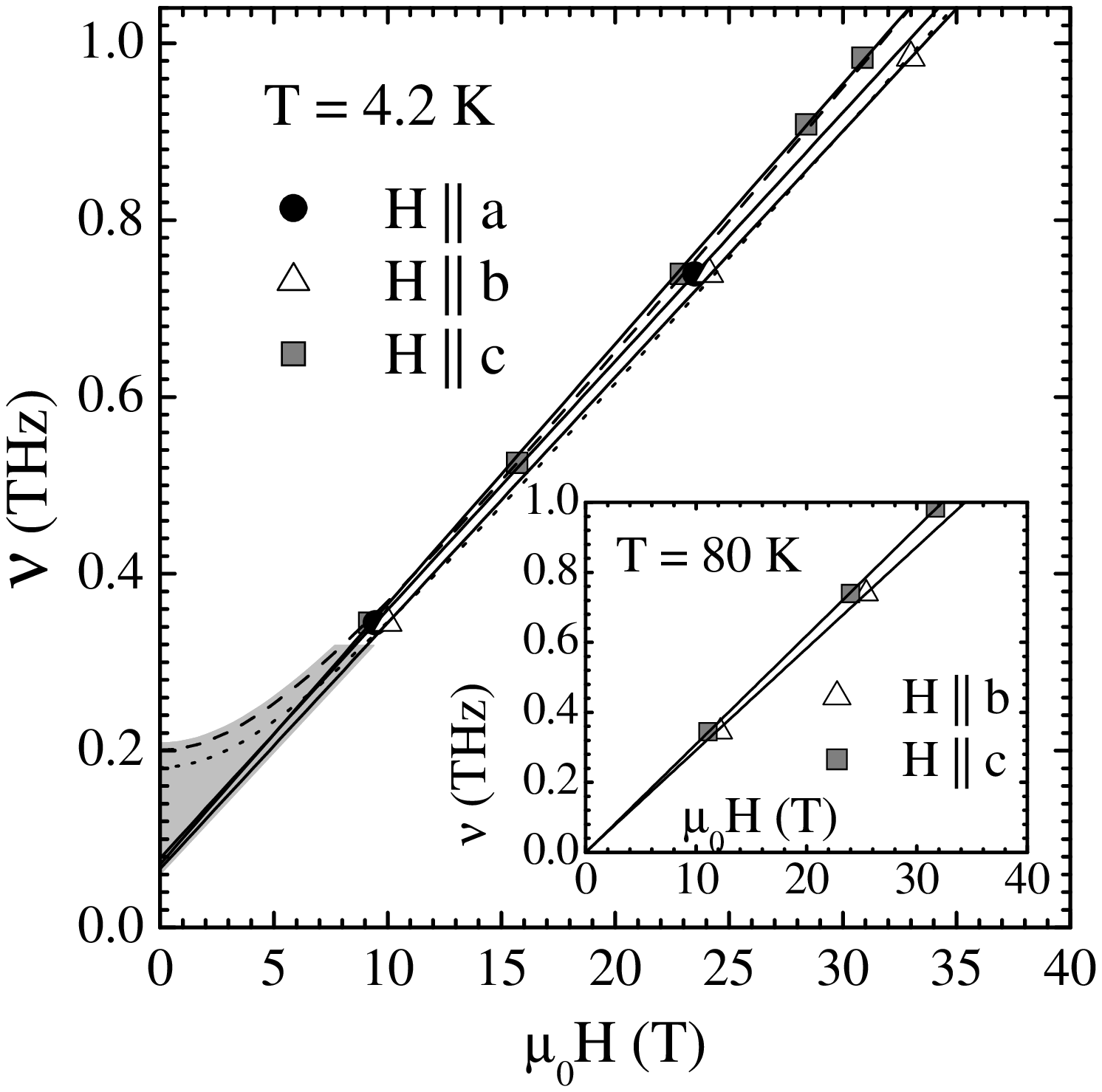}
\end{center}
\caption{Frequency $\nu$ vs. resonance field $H_{res}$ diagram for different directions of the magnetic field at 4.2 K (main panel) and 80 K (inset),
respectively. {The solid lines are linear fits to the experimental data. The dashed and dotted lines represent a square root dependence
$\nu\!=\!\sqrt{\Delta^2 + (g\mu_B/h)^2H^2}$. The shaded area indicates an uncertainty in the determination of the gap.} (See the text)}
\label{branches}
\end{figure}

\section{Discussion}

Obviously, considerable changes of ESR in the low-temperature regime are related to the AF order of CaCu$_2$O$_3$. The $T$ dependence of the static
magnetic susceptibility $\chi(T)$ and the field dependence of the magnetization $M(H)$ at $T\!=\! 4.2$ K of the studied crystal are shown in the
inset of figure~\ref{shift}. Consistently with the results of Kiryukhin {\it et al.} \cite{Kiryukhin01} the $\chi(T)$ dependence reveals a
characteristic N\'eel peak at $T_N\!=\!25$ K signalling an AF phase transition, whereas a nonlinear behaviour of $M(H)$ for $H\!\parallel\!a$ axis
can be attributed to a spin-flop transition at $\mu_0H_{sf}\!\sim\!3$ T \cite{Kiryukhin01}. In fact, the condition for the occurrence of the magnetic
ground state in coupled spin chains derived recently by Sengupta {\it et al.} \cite{Sengupta04}, {$J_r/J_b\!\leq\! 2.53\!\sqrt{J_c/J_b},$ which
relates the strength of the intra-chain exchange $J_b\,\sim\,2000$\,K, the inter-plane exchange $J_c\,\sim\,100$\,K and the exchange coupling in the
rungs $J_r\,\sim\,J_c$ \cite{Drechsler04,Kiryukhin01}}, respectively,  is easily satisfied in CaCu$_2$O$_3$ using the estimate $J_b/J_c\!\sim\!20$
obtained from the experimentally observed magnetic moment {$\mu\!=\!0.2\mu_B$} \cite{Drechsler04,Kiryukhin01}. Remarkably, the onset of the decrease
of $H_{res}$ as well as the kink in the dependence $\Delta H(T)$ at $\nu\!=\!344$ GHz coincides with the AF phase transition at 25 K. Observation of
the frequency offset $\Delta$ gives evidence for the opening of the {gap for magnon} excitations below $T_N$. In fact, the simple linear
interpolation of the data points in figure~\ref{branches} sets the smallest possible gap because the $\nu(H)$ dependence may saturate at low fields.
{E.g. in a simple case of a collinear "easy axis" or an "easy plane" antiferromaget one expects for the "hard" direction a square root dependence of
$\nu$ on $H$ in a form $\nu\!\simeq\!\sqrt{\Delta^2 + (g\mu_B/h)^2H^2}$ \cite{Turov}. Applying this relation to experimental data yields a larger
value of the gap $\Delta\!\simeq\!200$\,GHz (figure~\ref{branches}) corresponding to 0.8\,meV, or 10\,K. From our data it is not possible to
discriminate between these two behaviours which brings an uncertainty in the determination of the gap indicated by a shadowed area in
figure~\ref{branches}.} The assignment of the small spin gap to the host or to the minority spin subsystem is difficult at present, since the coupled
subsystems may exhibit rather different gaps in each subsystem despite the identical critical temperature. {In the absence of the long-range order
a}{ small spin gap may arise if spin chains are cut by structural imperfections in segments of a finite length. In this scenario ''extra'' spin
states {in a concentration $x$} may reside at the spin chain boundaries and} one would expect the corresponding impurity induced finite size gap of
the order of $\Delta_L=0.5\pi^2xJ_b\sim x10^4$~[K] \cite{Asakawa98}. {Such a singlet/triplet gap can be directly observed in} {ESR \cite{Sakai00} in
the presence of the alternating $g$~tensor and the anisotropic Dzyaloshinsky-Moriya (DM) interaction \cite{DM}}. However, taking the values of the
intra-chain AF coupling $J_b\!\sim\!2000$ K and the gap {$\Delta\!=\!3.4\,-\,10$\,K} one would arrive at an extremely low concentration of defects
{$x\!=\!\Delta_L/0.5\pi^2J_b\!\lesssim\!10^{-3}$} whereas the static susceptibility data yield much larger value of $x$ in the range of a few
percent. {In this case a gap of the order of 100~K should be expected {but which is not observed}.} Hence, a simple interruption of chains as the
cause of the observed spin gap in CaCu$_2$O$_3$ is unlikely. This suggests that the defect related minority spins should reside outside the zigzag
chains/pseudo-ladders {\cite{Hess}}. Such an appropriate position would be given e.g.\ by {some} Cu$^{2+}$ ions substituted for Ca.

Indeed, in a detailed structure analysis by Ruck {\it et al.} \cite{Ruck01} a sizeable amount of excess Cu $\sim\,10$\,\% (up to 14\%) and some
{oxygen O(2) deficiency on a pseudo-ladder rung of about 2 \% has been found {by} analyzing electron density measurements (Fourier synthesis with
x-ray data). \vl{Similar appreciable deviations from the stoichiometry are found in our crystals too (see Section~\ref{experiment})}. Using the
simple concept of bond length considerations {for the Cu valency $\nu_{Cu}$ versus bond lengths $s_{Cu-O}$} by Brown, Altermatt and O' Keeffe
\cite{Brown85}
\begin{equation}
\nu_{Cu}=\sum s_{Cu-O}=\sum \exp\left[(r_0-r)/B\right], \label{bondvalency}
\end{equation}
{where $B\,=0.37\,\AA$ is an empirical factor}, $r_0$(Cu$^{2+}$)=1.679 \AA \ and $r_0$(Cu$^{1+}$)=1.593 \AA\ { are the ionic radii of Cu in a 2+ and
1+ oxidation states, respectively,} Ruck~{\it et al.} \cite{Ruck01} ascribed the excess Cu solely to {nonmagnetic} Cu$^{1+}$ residing close to a Ca
position. However, near an O(2) vacancy such a position becomes unstable since the Cu-ion is now attracted by the two O(2)-ions belonging to a {
regular, i.e. oxygen non-deficient,} rung of an adjacent pseudo-ladder. As a result the corresponding small part of those Cu ions near an O(2)
vacancy is likely to be shifted towards that nearest regular rung, i.e.\ they will occupy real interstitial positions. This situation is illustrated
in figure~\ref{bondpic}, where {in the left panel}  the formal valency of a Cu ion in dependence on its position along the $c$~axis between two
neighbouring rungs according to Eq.~(\ref{bondvalency}) is depicted. In fact, a self-consistent Cu$^{2+}$ position occurs { about 1.5 \AA} above the
regular rung, i.e.\ {\it inside} a buckled pseudo-ladder far from the self-consistent Cu$^{1+}$ position  about 0.55 \AA \ below the opposite regular
rung {\it outside} a buckled pseudo-ladder. {The different positions of these ''defect'' Cu$^{1+}$ and Cu$^{2+}$ sites with respect to the ladders
are sketched in the right panel of figure~\ref{bondpic}. In particular, the position of the off-ladder Cu$^{2+}$ ion allows for a coupling $J_{eh}$
of its spin with the Cu spins at the regular lattice sites to the left and to the right, respectively, along the $a$ axis (see below). We note} that
such a small amount of ''defect'' Cu$^{2+}$ ions is in {agreement} with the small number of ''extra'' spins as deduced from the magnetic
susceptibility measurements.}
\begin{figure}
\begin{center}
\includegraphics[angle=-90,width=\columnwidth,clip]{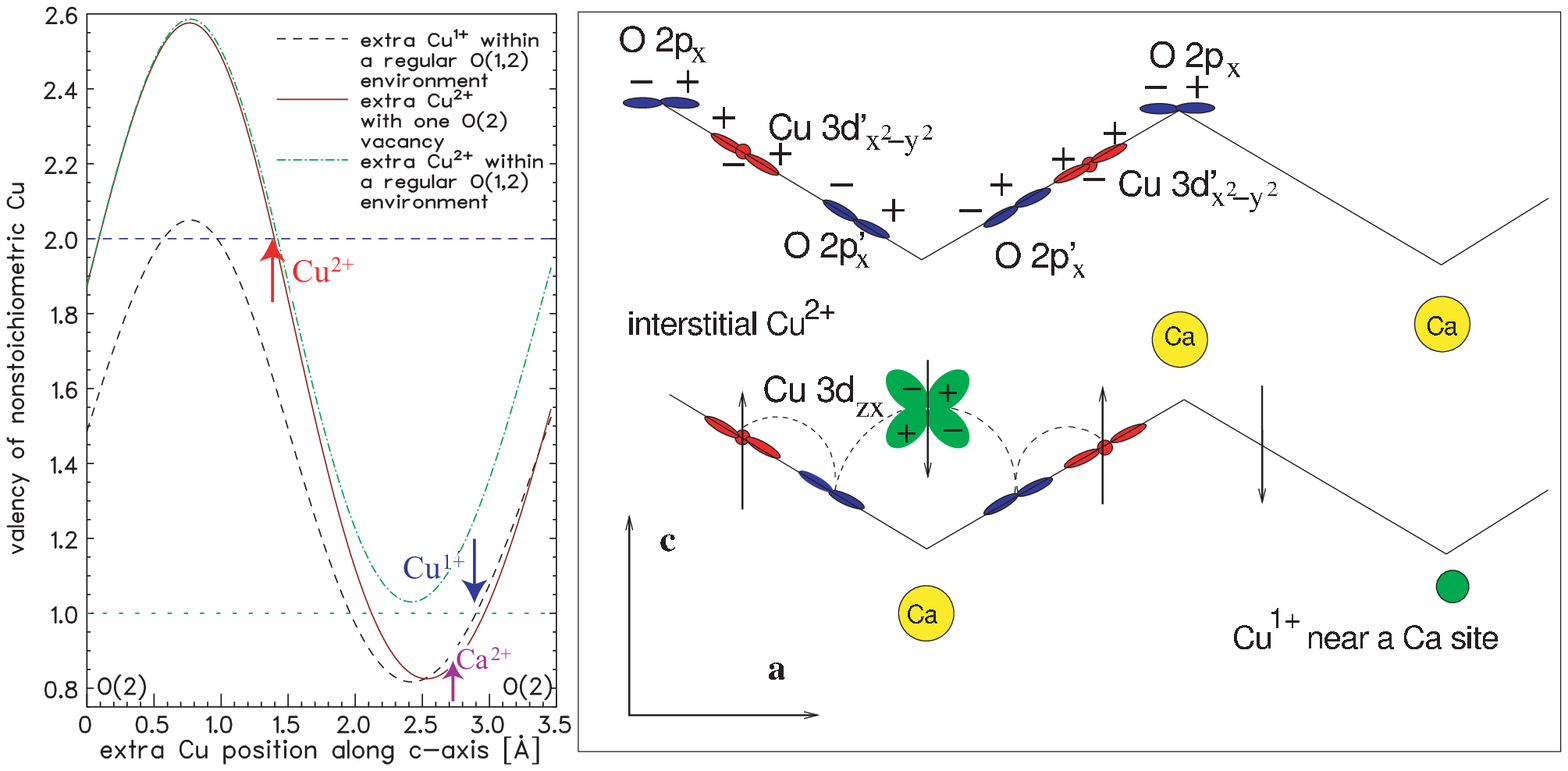}
\end{center}
\caption{{ Left panel: Cu valency vs.\ position along the $c$~axis in between two adjacent rung centres according to
Eq.~(\ref{bondvalency}). {Equilibrium positions of the ''defect'' Cu$^{1+}$ and Cu$^{2+}$ sites as well as of the Ca$^{2+}$ regular site are
indicated in the plot }{ adopting $r_0$ for Cu$^{1+}$ and Cu$^{2+}$, respectively, (see Eq.~(\ref{bondvalency})).} Right panel: The main orbitals
involved in the exchange interaction between an interstitial Cu$^{2+}$ {ion} and the {regular Cu$^{2+}$ ions in} the buckled pseudo-ladder structure
of CaCu$_2$O$_3$. The notation of orbitals with a prime corresponds to the local system of coordinates adopted to the zigzag (double) CuO$_2$ chains
rotated alternatingly around the $b$~axis. The signs $\pm$ stand for the phases of the $p$ and $d$ orbitals. {The Cu$^{1+}$ ion positions are shifted
closer to the buckled rung lines as compared to the Ca position (cf. left panel). Note the downshift of the interstitial Cu$^{2+}$ ion which
facilitates the coupling of its spin with the Cu spins at the regular pseudo-ladder sites.}} {The relevant superexchange paths are indicated by
dashed lines.}} \label{bondpic}
\end{figure}

Anyhow, the occurrence of a magnon gap in {an AF ordered} $s\!=\!1/2$ magnet is usually related with additional anisotropic corrections to the main
isotropic Heisenberg exchange. In case of CaCu$_2$O$_3$ it may be {the DM} interaction \cite{DM} which is, in principle, allowed in this compound
owing to the missing inversion symmetry centre between the neighbouring Cu ions (figure~\ref{structure}). {Moreover, the assumption of the DM
interaction between the spins in the ladder rung naturally explains} {details of the} the incommensurate spin structure in CaCu$_2$O$_3$
\cite{Kiryukhin01}. In fact, the observed noncollinearity of the rung magnetic moments defined by the angle $\eta $ (see
figure~\ref{magneticstructure}) as well as the related deviation from the antiferromagnetic orientation, measured by the angle $\Theta_r=\pi-\eta $,
is directly { connected} to the presence of DM interactions {\bf D} $\parallel$ {\bf b}. In classical spin and nearest neighbour approximations it
simply reads $\tan \Theta_r = D/J_r \approx 0.4$, where the experimental value $\eta \approx $ 160$^\circ $, from the neutron scattering data
\cite{Kiryukhin01}, has been used. Adopting $J_r \sim $ 10 meV \cite{Kiryukhin01,Drechsler04} a sizeable DM exchange $D \sim$ 4 meV would be
expected. Comparing this value with that for the well-known 2D cuprate La$_2$CuO$_4$ of 5 meV {\cite{Keimer93} or of about 1.5 meV for the
edge-shared chain compound Li$_2$CuO$_2$ \cite{Boehm98} i.e.\ all } values being {significantly} larger {than the gap derived from the ESR data}, one
might conclude that in our case the spin gap could be ascribed to the {coupling between the minority and majority spin subsystems. (We note that the
single-ion anisotropy is absent in the $S\,=\,1/2$ case.) Indeed the} { position of a Cu$^{2+}$ ion at a low-symmetry interstitial} site allows also
the occurrence of DM interaction terms for the ''extra'' spin\,-\,host exchange in addition to the isotropic symmetric exchange $J_{eh}$ shown in
figure\ \ref{magneticstructure} { (see below).} { The small value of the spin gap found in the present work is in {agreement} with recent inelastic
neutron scattering data \cite{Lake05}}. {No gap exceeding the experimental resolution limit of 3\,meV has been found, which sets an upper bound for a
possible spin gap in CaCu$_2$O$_3$.} {The possible upper bound can be further reduced to $\sim\,0.8$\,meV based on our ESR analysis
(figure~\ref{branches}). Moreover it yields also a lower bound of about 0.3\,meV.}

The magnitude of the anisotropy gap in CaCu$_2$O$_3$ amounting to {less than 10\,K} sets the respective field scale of a few Tesla for the
{spin-flop} transition {$H_{sf}\!=\!\Delta/g\mu_B$ \cite{Turov}}, which indeed occurs at $\mu_0H_{sf} \!\sim\!3$ T (see figure~\ref{shift} and
Ref.~\cite{Kiryukhin01}). {At fields $H\!>\!H_{sf}$} one expects a reduction of the staggered field and {saturation} of the {uniform} spin
susceptibility {$\chi_s$} below $T_N$.  This effect is particulary appreciable in strong magnetic fields yielding the decrease of the resonance shift
(Figs.~\ref{parameters_c}c and \ref{shift}). Furthermore, the $T$ dependence of the resonance linewidth $\Delta H(T)$ measured at 23 and 32\,T
changes qualitatively as compared to ESR in the "small" field of 11\,T giving a hint for possible recovery of the paramagnetic state in strong
fields. In low-dimensional spin systems the slowing down of the short-range AF fluctuations sets in at temperatures far above $T_N$ which results in
a continuous growth of the ESR width by approaching $T_N$ \cite{Benner}. This kind of behaviour is seen in the $\Delta H(T)$ dependence at the lowest
resonance field of $\sim\!11$ T (figure~\ref{parameters_c}b). However, at stronger magnetic fields of 24 and 32 T the signal {\it narrows}
continuously in the whole temperature range giving evidence for a strong suppression of AF correlations in CaCu$_2$O$_3$ by magnetic field. {
Theoretical calculations \cite{Chakr90}} { show that in a low-dimensional antiferromagnet the ESR critical broadening scales with the
spin-correlation length $\xi$ which implies that in CaCu$_2$O$_3$ $\xi$ reduces significantly in strong magnetic fields}. These ESR findings
corroborate the results of elastic neutron scattering in Ref.~\cite{Kiryukhin01} {which yield a finite $\xi$ of the order of 250 {\AA} already in a
moderate magnetic field of 8 T indicating the loss of the long-range AF order.}

\begin{figure}
\begin{center}
\includegraphics[angle=-90,width=0.95\columnwidth] {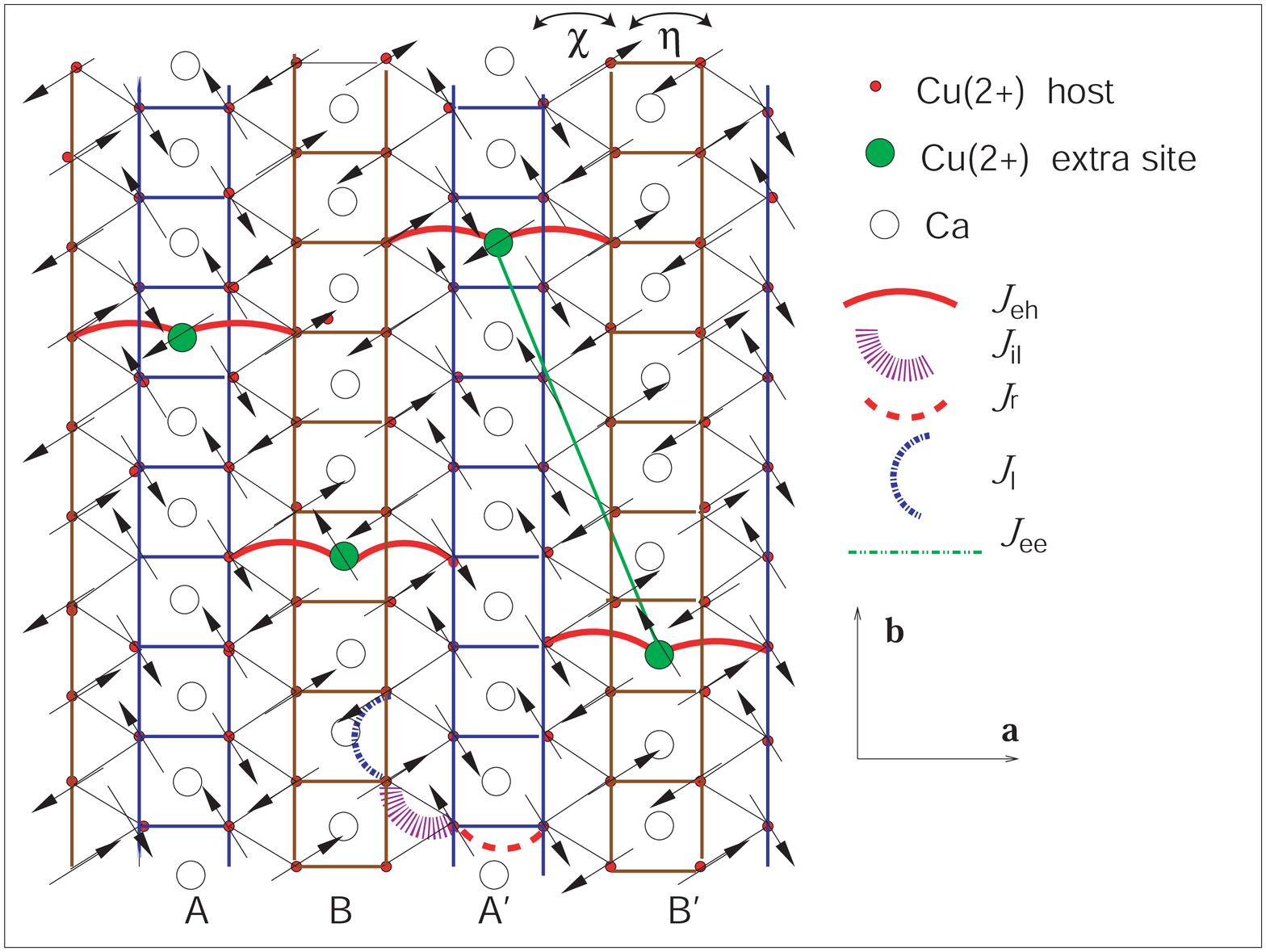}
\end{center}
\caption{{ Schematic magnetic structure of a buckled plane of pseudo-ladders in CaCu$_2$O$_3$ projected on the $(ab)$ plane with the most
important exchange interactions according to Eq.~(\ref{hierarchy}). Intra-ladder exchange along the legs and rungs, $J_l$ and $J_r$,
 is sketched by red-dashed and blue arcs, respectively. Frustrated inter-ladder exchange $J_{il}$ is shown by magenta arc. Much weaker exchange
 paths connecting
 ''extra'' spins at the Cu$^{2+}$ out-off ladder positions with the regular spin lattice and the coupling between the ''extra'' spins are
 depicted by solid
 red arcs and a green dash-double-dotted line, respectively.
$\eta$ and $\chi$ designate the angles between adjacent spins on the rungs and between the two legs of neighboring ladders, respectively. Note, that
in the real structure these angles slightly deviate from the ideal values of $\pi$ and $\pi/2$, cf. Fig.~1 of Ref.~\cite{Kiryukhin01}. (see the text
for details)}} \label{magneticstructure}
\end{figure}

A strong sensitivity of the magnetic order in CaCu$_2$O$_3$ to magnetic field is quite surprising because the magnetic energy scale is much smaller
than the leading exchange couplings in the system. One may speculate that owing to the frustration of the magnetic exchange \cite{Kim03,Drechsler04}
the minority spin subsystem plays a decisive role for the stabilization of the long-range N\'eel state. In particular, if the ''extra'' spins reside
at interstitial positions,  as discussed above, they are likely to mediate the second neighbour inter-ladder interaction in the highly frustrated $a$
direction {(figure~\ref{bondpic} and \ref{magneticstructure})}. This {antiferromagnetic coupling $J_{eh}$} is responsible for the intermediate
commensurate neutron scattering magnetic $h$ peaks shown in Figs.~6 and 8 of Ref.~\cite{Kiryukhin01} and helpful to promote the 3D AF order. {It
arises from a similar superexchange process as that which creates the leading coupling along the legs $J_{l} \sim$ 2000 K but it is strongly reduced
by the buckling and by the out-off-pseudo-ladder position
 of the ''extra'' (interstitial) Cu$^{2+}$ site
%position along the $c$-axis from the buckled pseudo-ladder plane
(figure~\ref{bondpic}). Applying standard 4th order perturbation theory within a {\it pd}-multiband extended Hubbard model (see e.g.
Ref.~\cite{Eskes93}) and the angular dependences for the Slater-Koster integrals from Refs.~\cite{Harisson80,Sharma79}, we may estimate
$J_{eh}\approx \sin ^2\left(\Theta '\right)\sin ^2\left(0.5(\Theta_r -\Theta ')\right) fJ_l \sim\,8\,-\,30$\,K, where $\Theta_r$ and $\Theta '$
denote the buckling angle and the O(1)\,-\,''extra'' Cu$^{2+}$ site\,-\,O(1) angle,} respectively (see above). Here $f\,\lesssim$\,1 is a further
reduction factor due to the larger distance between O(1) and the mentioned off ladder position of the ''extra'' spins compared with the usual Cu-O
distance of about 1.9 \AA \ in planar CuO$_4$ plaquettes. Here a Cu 3$d_{zx}$ or a mixed 3$d_{zx}$-3$d_{yz}$ state for the ''extra'' spins has been
adopted (figure~\ref{bondpic}). Noteworthy, the estimated above ''extra'' spin\,-\,host spin exchange $J_{eh}$ is of the same order as the ordering
temperature $T_N$ {suggesting that indeed the ''extra'' spin states are relevant for the long-range AF order and might also help developing the
incommensurate spin structure. Remarkably, the} weak commensurate correlation has been observed in the neutron-diffraction scans up to the highest
applied field of 8 T whereas the incommensurate one is shifted and/or suppressed (see figure\ 8 of Ref.~\cite{Kiryukhin01}). {Owing to a small
concentration of ''extra'' spins the coupling between them, most probably of the dipolar character, is expected to be very small $J_{ee}\, <$\,1\,mK
and can be neglected.} Thus, we arrive finally at an hierarchy of exchange integrals:
\begin{equation}
J_l > J_r \sim  \mid J_{il} \mid \gg J_{eh} \gg \mid J_{ee} \mid ,
\label{hierarchy}
\end{equation}
{ where $J_{il}$ denotes the ferromagnetic interladder exchange} {which is of the order of the rung exchange $J_r$}. {As the last two exchange
integrals in Eq.~(\ref{hierarchy}), $J_{eh}$ and $J_{ee}$, are relatively weak and from the remaining frustration in the nearest neighbour
inter-ladder coupling, one indeed may expect a strong influence already of a moderate field on the magnetic order. The relevant interaction paths
yielding the spin structure of CaCu$_2$O$_3$ are shown in figure~\ref{magneticstructure}. The drawn spin lattice has been simplified for clarity: (i)
To emphasize that the observed commensurate component in the elastic structure factor (Ref.~\cite{Kiryukhin01}) is caused by the AF ''extra''
spin\,-\,host spin interaction $J_{eh}$ a slight incommensurability along the $a$ axis ([0.429,0.5,0.5]) due to the frustrating inter-ladder exchange
$J_{il}$ within local triangles of the Cu host spins has been ignored; (ii) In the real structure both the depicted antiparallel spin arrangement on
the rungs and the orthogonal orientation at the inter-ladder borders are slightly distorted; (iii) Due to the presence of DM interactions for
neighbouring spins at the buckled rungs, the magnetic moments are not in the ($ab$) plane, but within the ($ac$) plane. (iv) The pseudo-ladders A(A')
and B(B'), respectively, being relatively shifted by $0.5b$ along the $b$ axis are alternatingly buckled (cf. figure~\ref{structure}). }

{Because of the relative smallness of the ''extra'' spin - host spin exchange $J_{eh}$ as compared to other exchange integrals the relaxation of
''extra'' spins can be treated in a close analogy to nuclear magnetic resonance where the nuclear spin relaxation rate $1/T_1$ is determined by a
fluctuating hyperfine field of electron spins \cite{Moriya63}. Following this analogy the $T_1$-contribution to the ESR linewidth of ''extra'' spins
$\Delta H\sim 1/\gamma T_1$ is given by the dynamical susceptibility $\chi(\omega)$ of the host spins $1/T_1T\sim
J_{eh}\mathrm{Im}(\chi(\omega))/\omega$ \cite{Kochelaev94,Kataev97}. Here $\gamma$ is the gyromagnetic ratio. Since the AF correlation length $\xi$
is inversely related to the frequency of the fluctuations of the spins, the increase of $\xi$ will result in the slowing down of spin fluctuations,
i.e. to the shift of the spectral weight of the dynamical susceptibility of bulk spins to lower frequencies, comparable with the ESR excitation
frequency, and consequently to a strong broadening of the ESR signal of ''extra'' spins. Thus the} suppression of the long-range AF order and spin
fluctuations by a strong magnetic field may explain why the ESR signal is observable only at sub-Terahertz frequencies and not  in the low-frequency
domain of $10-100$ GHz. Plausibly the corresponding resonance field of at most 3~T is not sufficient to suppress AF fluctuations which broaden the
resonance line very strongly. Even at a much higher frequency of 344 GHz the width of the ESR signal approaches 3~T at low $T$. The detection of such
a broad resonance at smaller probing frequencies would be hardly possible \cite{remark_Schwenk}.

\section{Conclusions}

We have measured high field ESR of a small amount of "quasi-free" spins which determine the bulk static magnetic properties of a
$s\!=\!1/2$-pseudo-ladder compound CaCu$_2$O$_3$ at low temperatures. The frequency-, magnetic field- and temperature dependences of the ESR signal
show that these ''extra'' spin states are strongly coupled to the bulk spins and are involved in the AF ordering of the host spin lattice at
$T_N\!=\!25$ K. In particular, ESR gives evidence for the opening of a gap for magnetic excitations below $T_N$ whose small magnitude explains the
occurrence of a {spin-flop} transition in a small magnetic field of $\sim\!3$ T. Another remarkable observation is the narrowing of the ESR line {at
fields above 20~T} which signals the suppression of AF {correlations} and may indicate the loss of the long-range magnetic order in strong magnetic
fields. Experimental ESR data corroborate theoretical predictions of a strong interplay between ''extra'' spin states at the imperfections of the
host AF $s\!=\!1/2$-lattice with the bulk spins.

\section*{Acknowledgement}
The DFG under project SP 1073 (S-LD and BB) is gratefully acknowledged for financial support. Work of RK in Toulouse was supported by the DFG through
KL 1824/1-1. This work was also supported in part by the MRSEC Program  of the National Science Foundation under award number DMR 02-13282. We thank
C Hess, B Lake, G Krabbes, A Moskvin, K-H M\"uller and M Wolf for useful discussions.

%\end{multicols}
\end{document}